\documentclass[9pt,twocolumn,twoside]{revtex4-1}
\usepackage{bm}
\usepackage{csquotes}
\usepackage{mathtools}
\usepackage{textcomp}
\usepackage{amsmath}
\usepackage{amssymb}
\usepackage[utf8]{inputenc}
\usepackage{xcolor}
\usepackage[normalem]{ulem}
\usepackage{amsthm}
\usepackage{enumitem}
\theoremstyle{remark}
\DeclareFontFamily{U}{wncy}{}
\DeclareFontShape{U}{wncy}{m}{n}{<->wncyr10}{}
\DeclareSymbolFont{mcy}{U}{wncy}{m}{n}
\DeclareMathSymbol{\Sha}{\mathord}{mcy}{"58}
\usepackage[pretty]{revquantum}
\makeatletter
\renewcommand{\fnum@figure}{Fig.\thefigure}
\makeatother
\begin{document}
	\title{An Experimental Scheme for Testing Molecular Rectification with Only Micrometer-Scale Fabrication Requirements}
	
	\author{Jiantang Jiang}\thanks{Corresponding author: Jiantang Jiang (email: tumport@126.com)} 
	\affiliation{Zhongshan Shiruan Software Technology Co., Ltd., Zhongshan City, Guangdong Province 528429, China
	}
	\begin{abstract}
	In our previous work, we proposed a theoretical model capable of inducing sustained directed transport without consuming information or external energy. However, the experimental verification schemes proposed previously were technically challenging, and the model has therefore not yet been experimentally tested. In this work, we propose a greatly simplified experimental design. By introducing hydrophilic functional groups onto the tip of a gold needle through surface modification and pressing the needle against one of two liquid–vapor interfaces, the compressed interface is maintained at a higher ionic concentration than the other interface, thereby sustaining a concentration difference between them. This design reduces the fabrication requirement from the nanometer scale to the micrometer scale, substantially lowering experimental complexity and facilitating experimental examination of the proposed mechanism. 
	\end{abstract}
	
	\maketitle
	
	\section{Introduction}
	\label{sec:1}
	Since Maxwell’s seminal thought experiment\cite{01}, the rectification of thermal fluctuations has been intimately connected to information processing and energy consumption. Experimental realizations of Maxwell's demon–type feedback controllers have been demonstrated across a wide range of platforms\cite{02,03,04,05,06,07,08,09}. Extensive studies of biological\cite{10,11,12,13,14} and synthetic\cite{15,16,17,18,19,20,21,22} ratchets have reinforced this paradigm: asymmetric potentials or geometric constraints, in the absence of energy or information input, are insufficient to induce a nonzero probability current. Directed transport is therefore generally attributed to time-dependent driving, feedback control, or nonequilibrium chemical reactions. Whether purely static asymmetries can induce directed transport without invoking an external resource remains an open and fundamental question.
	
	In our previous work, we proposed a model capable of inducing sustained directed transport without consuming external energy or information\cite{23} (see brief introduction of the model in Section.~\ref{sec:1} of Supplementary Information). The model was subsequently refined and shown to be consistent with molecular dynamics simulations\cite{24} (see simulation results in Section.~\ref{sec:2} of Supplementary Information). Experimental verification of this model would have important implications for our understanding of nonequilibrium transport and the foundations of statistical physics. However, previous experimental designs required nanometer-scale fabrication precision, resulting in high cost and complexity that have hindered experimental testing. Here, we propose an experimental scheme requiring only micrometer-scale fabrication and derive quantitative relationships among the relevant control parameters. This design substantially reduces the cost and complexity of experimental verification and may facilitate direct testing of the proposed mechanism.
	
	\section{Methods}
	\label{sec:2}
\begin{figure}[t]
    \centering
    \includegraphics[width=0.7\columnwidth]{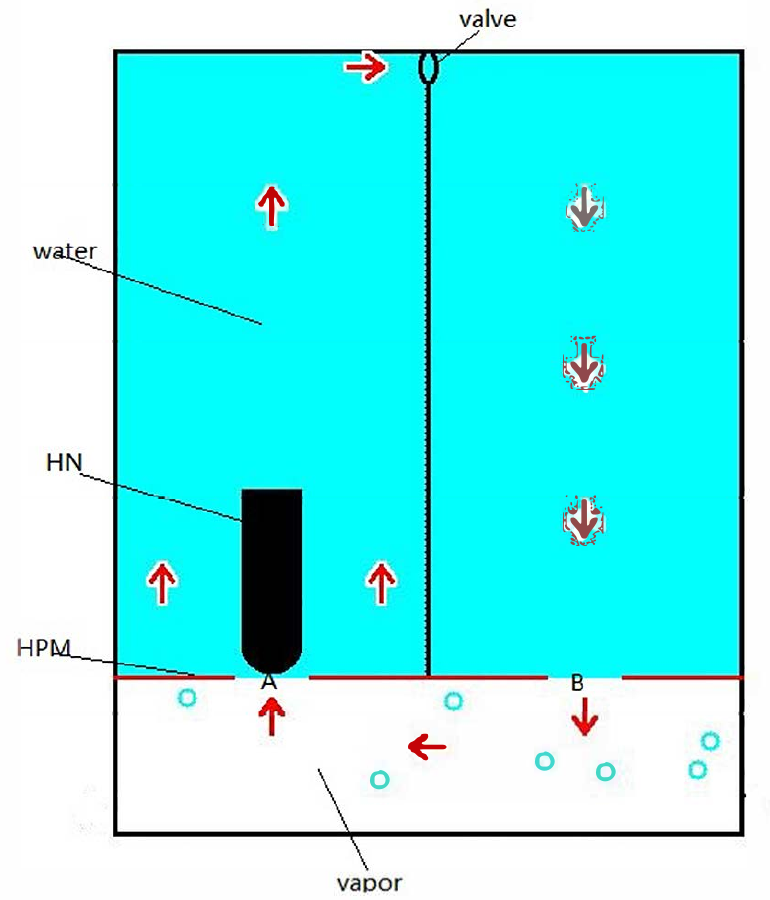}  
    \caption{Schematic illustration of the experimental system. HPM: hydrophobic porous membrane; HN: hydrophilic needle.}
	\label{fig:1}
    \end{figure}

	As illustrated in Fig.~\ref{fig:1}, the experimental system consists of a square container in which a hydrophobic porous membrane (HPM) separates liquid water from water vapor. A partition located at the center divides the liquid phase into left and right water columns, while a valve is installed at the top of the partition. Two pores in the HPM allow molecular exchange between the liquid columns and the vapor phase, thereby forming two liquid–vapor interfaces, denoted A and B. Owing to the hydrophobicity of the membrane, the liquid does not penetrate the pores within the pressure range permitted by capillary forces.

\begin{figure}[t]
    	\centering
    	\includegraphics[width=0.7\columnwidth]{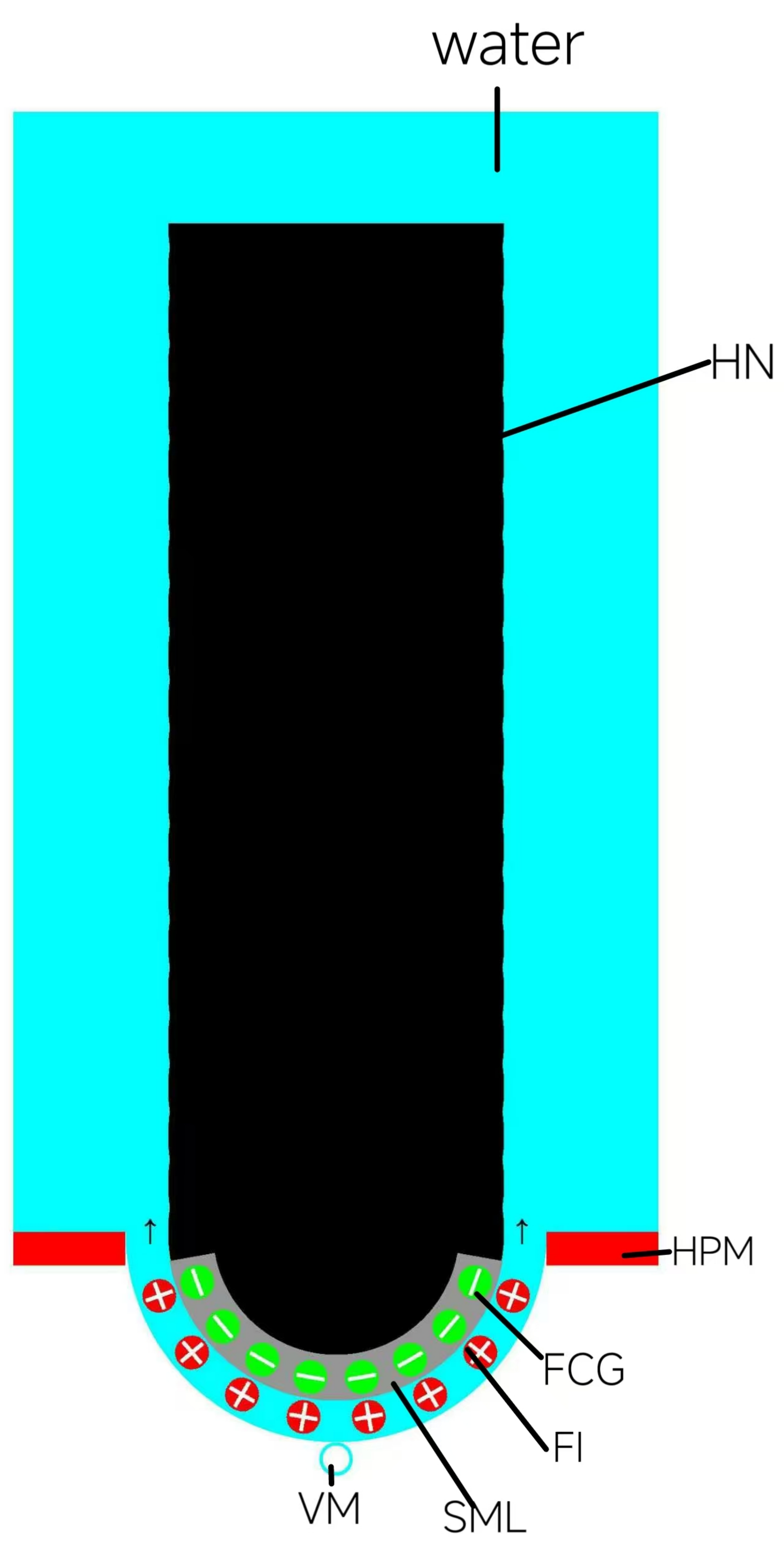}  
    	\caption{Interface A deformed by the needle tip. HPM: hydrophobic porous membrane; HN: hydrophilic needle; FCG: fixed charged groups; FI: free ions; SML: surface modification layer; VM: vapor molecule.}
    	\label{fig:2}
    \end{figure}

    A vertically movable microneedle is positioned above interface A. Through surface modification, the tip of the needle is rendered strongly hydrophilic. As illustrated in Fig.~\ref{fig:2}, the surface modification layer(SML) on the needle tip dissociates in water, producing fixed charged groups(FCG) and free ions(FI). When the needle tip presses against interface A, the ionic concentration near interface A becomes higher than that near interface B, thereby reducing the evaporation rate of interface A. If the molecular rectification model is valid, a sustained directed flow is expected to emerge in the system, as indicated by the arrows in Fig.~\ref{fig:1}.
    
    \section{Analysis and Discussion}
    \label{sec:3}
    The present experimental design is based on the same physical mechanism as the models described in Refs.\cite{23} and \cite{24}. The discussion below focuses primarily on experimental implementation, while the detailed dynamical principles can be found in the original references.
    
    We assume that water is used as the solvent, that the needle tip is hemispherical, and that its diameter is approximately equal to the pore diameter. Surface modification can be employed to form a layer of 3-mercaptopropyl sulfonate (MPS)\cite{26} on the tip of the gold needle. The thickness of this layer is only on the nanometer scale. As shown in Fig.~\ref{fig:2}, when the needle is immersed in water, the MPS layer dissociates and releases free cations. When the needle applies a normal pressure to interface A, the energy released by interfacial contraction is insufficient to overcome the interaction between the charged groups and the free ions, resulting in the formation of a wetting layer surrounding the needle tip.
    
    The pressure   (Pa) applied by the needle and the radius of curvature of the deformed interface are related through the Laplace equation:
    \begin{equation}
    	P_n=\frac{2 \sigma}{r_c}
    	\label{equ:1}
    \end{equation}
    where $\sigma$ is the surface tension of the liquid ($N/m$), and $r_c$ is the radius of curvature of the deformed interface ($m$).
    
    The wetting layer is compressed by $P_n$ and becomes thinner. As the layer becomes thinner, a larger ionic concentration difference develops between the wetting layer and the bulk liquid. Under isothermal conditions, equilibrium is reached when the chemical potential energy released by transferring water molecules from the bulk phase into the wetting layer is exactly balanced by the mechanical work associated with expansion of the wetting layer:
    \begin{equation}
    	P_nV_m = \mu_{conc,bulk} - \mu_{conc,a}
    	\label{equ:2}
    \end{equation}
    where $\mu_{conc,bulk}$ and $\mu_{conc,a}$ denote the concentration-dependent chemical potentials of water molecules in the bulk liquid and in the interfacial layer A, respectively, and $V_m$ is the molar volume of water ($m^3/mol$).
    
    In classical osmotic equilibrium, the role of a semipermeable membrane is simply to prevent the transport of solute particles while allowing solvent molecules to pass freely. Although no physical semipermeable membrane exists between the bulk liquid and the wetting layer in the present system, the fixed charged groups perform an analogous function by confining the free ions within the wetting layer while still permitting unrestricted transport of water molecules. Consequently, the bulk liquid and the wetting layer may be regarded as two phases separated by an effective ion-selective boundary, making their equilibrium condition mathematically analogous to that of classical osmotic equilibrium. The ionic concentration difference ($C_a - C_b$) between interfacial layers A and B is therefore equivalent to the concentration difference C ($mol/m^3$) corresponding to an osmotic pressure $\pi$ ($Pa$) across a semipermeable membrane. Assuming an ideal dilute solution, the van't Hoff equation applies:
    \begin{equation}
    	\pi = CRT
    	\label{equ:3}
    \end{equation}
    where $R$ is the ideal gas constant and $T$ is the absolute temperature ($K$).
    
    Replacing $\pi$ with $P_n$ and rearranging yields:
    \begin{equation}
    	C_a - C_b = \frac{P_n}{RT}
    	\label{equ:4}
    \end{equation}
    Combining Eqs.~\ref{equ:1} and ~\ref{equ:4} gives:
    \begin{equation}
    	C_a - C_b = \frac{2 \sigma}{r_c RT}
    	\label{equ:5}
    \end{equation}
    
    Pure water undergoes self-ionization, with an ionic concentration of approximately $10^{-7} mol/L$ at room temperature. In addition, a small number of free ions may occasionally escape from the influence of the charged groups and diffuse into the bulk liquid. However, both effects are negligible for the present analysis. Therefore, the ionic concentration at interface B can be approximated as zero relative to that at interface A; i.e. $C_b \approx 0$, and thus:
    \begin{equation}
    	C_a = \frac{2 \sigma}{r_c RT}
    	\label{equ:6}
    \end{equation}
    
    As long as $P_n > 0$, it follows that $C_a > C_b$, and their difference is proportional to $P_n$ under the assumptions of an ideal dilute solution. According to Raoult's law, the vapor pressure decreases with increasing solute concentration. Therefore, when $P_n > 0$, the resulting concentration difference between interfaces A and B generates a vapor-pressure difference, promoting condensation at interface A. The spontaneous contraction of the liquid surface then drives the condensed water through the wetting layer into the bulk liquid, as indicated by the arrows in Fig.~\ref{fig:2}.
    
    Because the charged groups in the MPS layer are fixed, electrostatic attraction keeps the free cations localized near the needle tip. Consequently, continuous condensation does not eliminate the concentration difference between interfaces A and B. The system can therefore sustain the directed transport cycle illustrated in Fig. 1: water evaporates from interface B, vapor migrates from B to A, condensation occurs at interface A, and the condensed liquid returns to interface B through the open valve, forming a closed circulation loop. Observation of a steady directed flow that does not decay with time would provide evidence consistent with the proposed model.
    
    If the valve is closed to block the circulation, a constant pressure difference $P_d$ will develop between the two sides of the partition when interface A reaches vapor–liquid equilibrium. This pressure difference increases the chemical potential of the water column on side A and thereby enhances evaporation at interface A, producing an effect opposite to that of $P_n$.
    
    According to the Gibbs fundamental equation, under isothermal conditions the chemical-potential change associated with a pressure difference $P$ is:
    \begin{equation}
    	\mu_{pressure} = V_mP
    	\label{equ:7}
    \end{equation}
    The equilibrium condition is reached when the effects of $P_n$ and $P_d$ on the chemical potential of water molecules at interface A exactly cancel each other:
    \begin{equation}
    	\mu_{pressure} - (\mu_{conc,bulk} - \mu_{conc,a}) = 0
    	\label{equ:8}
    \end{equation}
    Comparison of Eqs.~\ref{equ:2} and ~\ref{equ:8} shows that, at equilibrium: $P_d = P_n$.
    
    For example, taking water as the solvent with a surface tension of 0.072 $N/m$ at room temperature and a pore diameter of 1 $\mu m$, Eq. ~\ref{equ:1} gives a theoretical maximum pressure on interface A of $P_{max}=288,000 Pa$, corresponding approximately to the pressure exerted by a 29-m-high water column. If the needle applies a constant pressure of $P_n=P_{max}/2 = 144,000 Pa$, the equilibrium pressure difference will be $P_d = P_n = 144,000 Pa$.
    
    Note that $P_d$ and $P_n$ act in the same direction on interface A. Therefore, $P_n$ cannot exceed $P_{max}/2$ ; otherwise, continuous condensation would cause interface A to bulge outward until the needle eventually penetrates the pore, leading to collapse of the system.
    
    Even if the pore diameter is increased to 0.1 $mm$, the predicted pressure difference remains approximately 1.44 $kPa$, which can readily be measured using conventional instrumentation. This suggests that the fabrication requirement may even be relaxed to the millimeter scale.
    
    Several additional points should be noted:\\
    (1) The needle merely applies a constant force (or equivalently, a constant pressure) to the liquid interface—for example, by connecting its opposite end to a spring—and does not continuously perform mechanical work. This is consistent with the original model, in which sustained directed transport is induced without continuous consumption of external energy.\\
    (2) Although the pore is fabricated from a hydrophobic material such as polytetrafluoroethylene (PTFE), interactions between the pore wall and interface A may still affect the surface tension. This effect has been neglected in the present calculations.\\
    (3) The hemispherical needle tip is adopted only for convenience of discussion. In principle, the needle geometry is unrestricted. Any configuration satisfying   can generate a concentration difference between interfaces A and B.\\
    (4) In addition to gold, well-established surface-modification techniques are available for rendering glass, silica, and metal oxides strongly hydrophilic \cite{27,28}, making these materials equally suitable for the present experimental scheme. Furthermore, the proposed scheme is not limited to water as the solvent. In principle, any solvent may be used, provided that the modified needle tip interacts sufficiently strongly with the solvent molecules to reduce the chemical potential of the molecules in the interfacial layer A, thereby suppressing the evaporation rate of interface A. Under these conditions, the directed transport described above is expected to be observable.\\
    (5) If air is present in the vapor region, solvent vapor must diffuse through the air from interface B to interface A, reducing the transport rate. We therefore recommend evacuating the air and retaining only pure solvent vapor in the vapor region.\\
    (6) Gravity acts symmetrically on the two liquid columns and has therefore been neglected. However, gravity contributes to the pressure at the bottom of the liquid columns and must be considered when large pore sizes or high liquid levels are involved.
    
    Additional suggestions relevant to experimental implementation are provided in Section.~\ref{sec:3} of the Supplementary Information.
    
    \section{Relationship Between Needle Pressure and Vapor Pressure}
    \label{sec:4}
\begin{figure}[t]
    	\centering
    	\includegraphics[width=0.7\columnwidth]{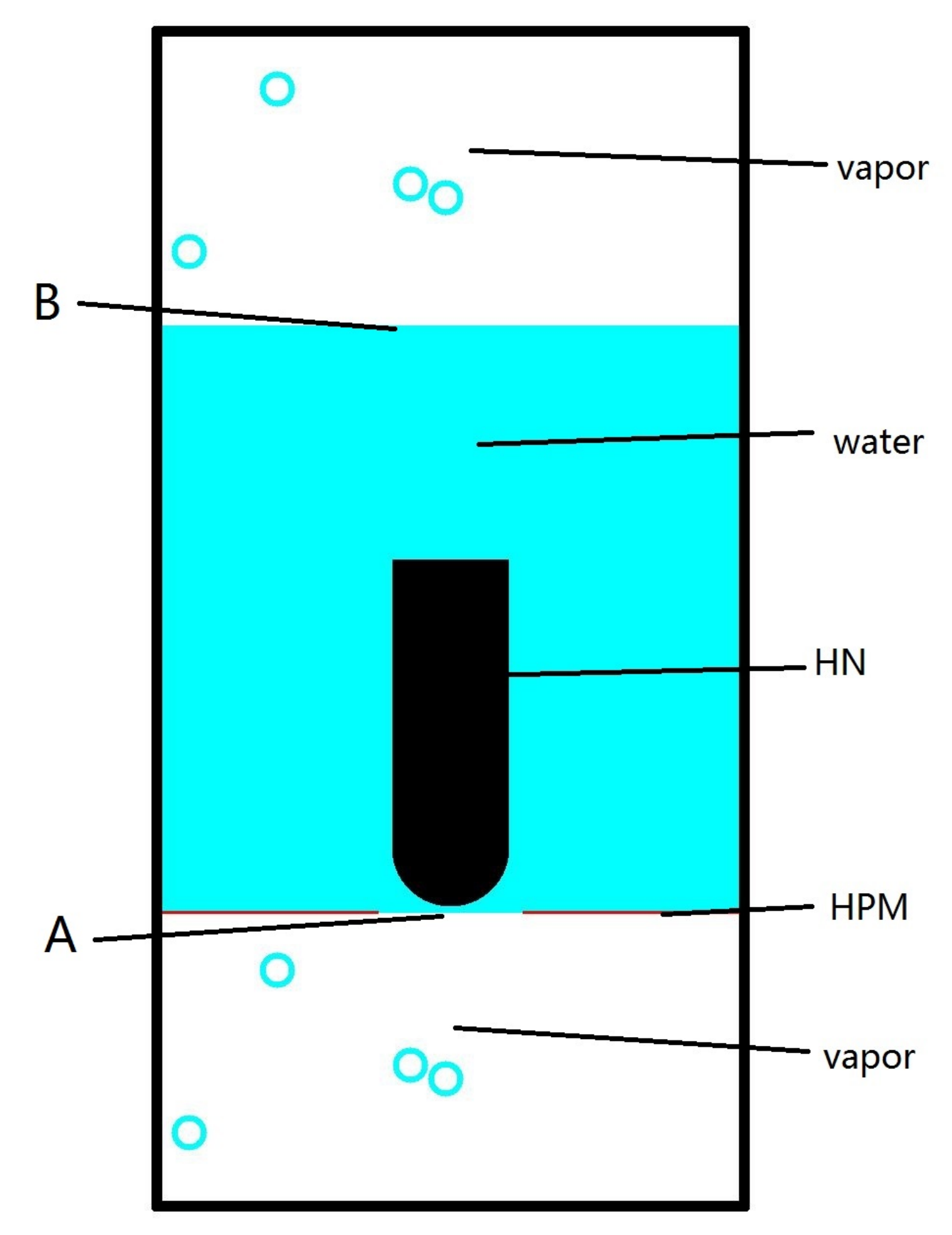}  
    	\caption{Schematic illustration of the apparatus used to measure the effect of needle pressure on vapor pressure. HPM: hydrophobic porous membrane; HN: hydrophilic needle.}
    	\label{fig:3}
    \end{figure}

    As illustrated in Fig.~\ref{fig:3}, the apparatus contains upper and lower vapor regions. According to the dynamical principles of the model, at a given temperature $T$ the saturated vapor pressure in the upper vapor region, $P_b$ , is equal to the vapor pressure of the pure solvent, whereas the saturated vapor pressure in the lower vapor region, $P_a$ , varies with the pressure $P_n$ applied by the needle to interface A. Treating the vapor as an ideal gas:
    \begin{equation}
    	P_a = P_b \exp(\mu_c / (RT))
    	\label{equ:9}
    \end{equation}
    where $R$ is the gas constant, $T$ is the absolute temperature ($K$), and $\mu_c$ is the chemical-potential difference between the two vapors phases.
    
    Since the upper vapor region is in equilibrium with pure water, its chemical potential is identical to that of bulk pure water. Therefore, $\mu_c$ originates entirely from the change in the chemical potential of water molecules of interfacial layer A induced by the applied needle pressure $P_n$. From Eq.~\ref{equ:2}, the relationship between $P_n$ and $\mu_c$ can be written as:
    \begin{equation}
    	\mu_c = -P_n V_m
    	\label{equ:10}
    \end{equation}
    where the negative sign indicates that compression by the needle lowers the chemical potential of water molecules within the wetting layer.
    
    Substituting Eq.~\ref{equ:10} into Eq.~\ref{equ:9} yields:
    \begin{equation}
    	P_a = P_b \exp(-P_n V_m/(RT))
    	\label{equ:11}
    \end{equation}
    
    For example, at $T = 300 K$, the saturated vapor pressure of pure water is approximately $P_b = 3550 Pa$. Taking $P_n = 288,000 Pa$, corresponding to the internal pressure of a $1-\mu m$ water droplet, Eq.~\ref{equ:11} gives $P_a = 3542.6 Pa$. The vapor-pressure difference between the upper and lower vapor regions is therefore only $7.4 Pa$, corresponding to less than one ten-thousandth of atmospheric pressure. Experimental verification would thus require high-precision instrumentation.
    
    Substituting Eq.~\ref{equ:1} into Eq.~\ref{equ:11} gives:
    \begin{equation}
    	P_a = P_b \exp[-(2\sigma V_m)/(r_c RT)]
    	\label{equ:12}
    \end{equation}
    
    Equation~\ref{equ:12} has the same mathematical form as the Kelvin equation describing the vapor pressure of liquid droplets, although the underlying physical mechanism is different. Section.~\ref{sec:4} of the Supplementary Information provides estimates of the flow rates driven by the saturated vapor pressure difference between interfaces A and B for different pore radii.

	\section{Conclusions}
	\label{sec:5}
	The conversion of random thermal motion into directed transport is a fascinating topic that may motivate a reexamination of some fundamental concepts in thermodynamics. The experimental design proposed here substantially lowers the technical barriers to testing the model and may facilitate experimental investigation of the proposed mechanism.
	
	\section*{Data availability}
	The data are available from the corresponding author on reasonable request.
	
	The author declares no competing financial or non-financial interests. 
	
	Correspondence should be addressed to Jiantang Jiang (tumport@126.com).

\end{document}


\title{Supplementary Information for the manuscript "An Experimental Scheme for Testing Molecular Rectification with Only Micrometer-Scale Fabrication Requirements"}

\maketitle

\section{The Model Proposed in Ref.\cite{S1}}\label{supp_sec:1}
Ref.\cite{S1} proposed a model schematically illustrated in Supplementary Fig.~\ref{fig:s1}. The system consists of a sealed P-shaped container partially filled with an extremely dilute electrolyte solution (blue), generating two liquid surfaces, A and B. Both liquid surfaces are in contact with the same vapor region above them. A hydrophobic porous membrane (HPM) is positioned directly above liquid surface A, allowing vapor molecules to migrate through its tiny pores. Two external electrodes (UEE and LEE) are placed above and below the left half of the container, respectively.

\begin{figure}[H]
    \centering
    \includegraphics[width = 0.35\linewidth]{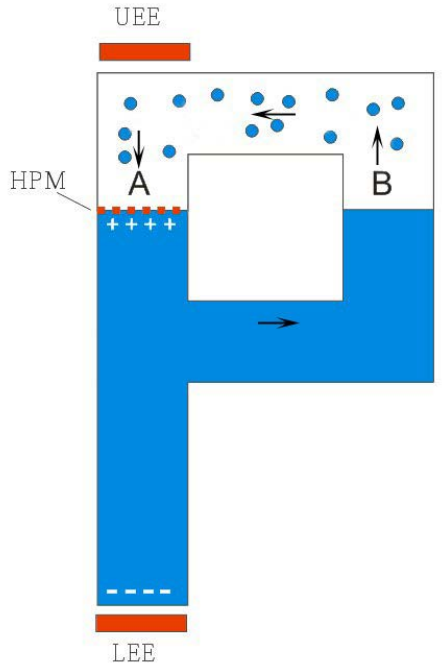}
    \caption{Side view of a molecular rectification mechanism. HPM denotes the hydrophobic porous membrane; UEE and LEE denote the upper and lower external electrodes, respectively.
    }
    \label{fig:s1}
\end{figure}

According to Ref.\cite{S1}, application of an external electric field by the two electrodes induces a sustained solvent flux in the system, as indicated by the arrows in Supplementary Fig.~\ref{fig:s1}.

Interestingly, Ref.\cite{S1} pointed out that the external electric field performs no continuous work. Instead, the energy driving the sustained directed transport originates from the work associated with interfacial contraction during condensation. As illustrated in Supplementary Fig.~\ref{fig:s2}, when a vapor molecule first contacts the liquid surface (arc op), the interfacial area associated with the molecule (yellow edge) is relatively large. The surface subsequently contracts spontaneously to minimize surface energy, thereby pushing the molecule into the liquid bulk. The energy released during this interfacial contraction was argued to provide the driving force for both the liquid flow and the vapor flow.
\begin{figure}[H]
	\centering
	\includegraphics[width = 0.35\linewidth]{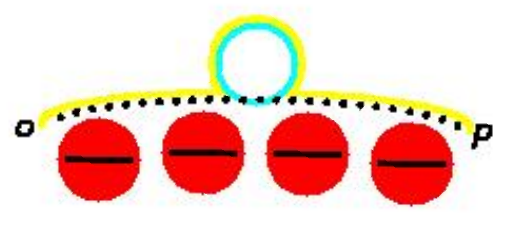}
	\caption{Condensation and surface contraction at interface A.
	}
	\label{fig:s2}
\end{figure}

Our previous study\cite{S2} demonstrated through molecular dynamics simulations that the above model can indeed induce sustained directed transport. However, it also revealed a significant challenge: weak electric fields produce only extremely weak transport, whereas strong electric fields can trigger ion escape, which ultimately leads to system collapse unless the free ions possess sufficiently large masses.

\section{Simulation of the Optimized Model}\label{supp_sec:2}
To prevent free ions from escaping, an optimized model was proposed in Ref.\cite{24}, and molecular dynamics simulations were performed to investigate it.

The simulation setup is shown schematically in Supplementary Fig.~\ref{fig:s3}. A simulation box (dimensions: $l_x = l_y = 2.6 nm$ and $l_z = 13 nm$) contains a 4-nm-thick water slab (950 molecules) centered along the z direction. A 1-nm-thick hydrophobic porous membrane (HPM) with a central pore of diameter 1.6 nm is placed on top of the slab and held fixed. Four fixed charges, all lying in the same xy plane, are attached to the pore sidewall. Four free ions of opposite charge disperse in the water, ensuring overall charge neutrality. Two liquid–vapor interfaces are formed: interface A at the pore opening and interface B at the free lower surface.
\begin{figure}[H]
	\centering
	\includegraphics[width = 0.35\linewidth]{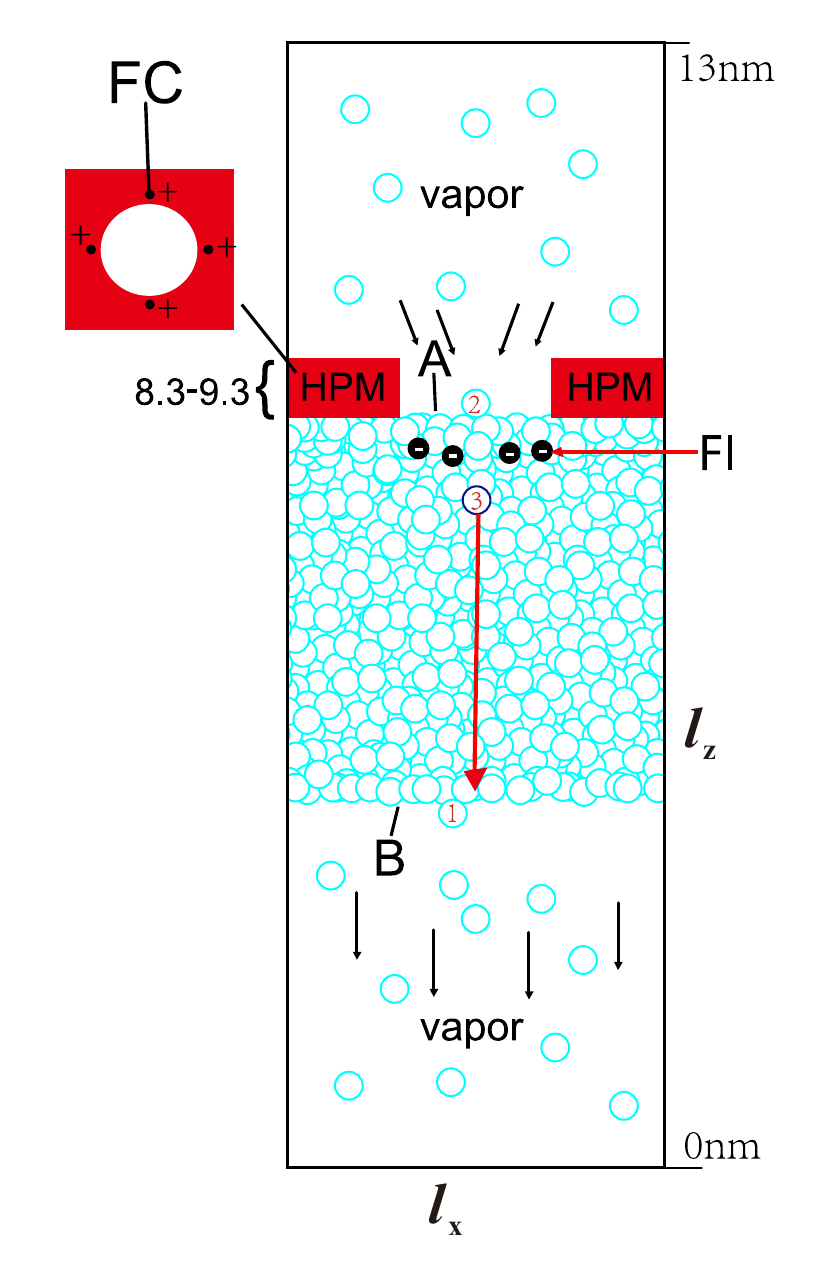}
	\caption{Simulation setup. HPM, hydrophobic porous membrane; FC, fixed charges attached to the pore sidewall; FI, free ions in water. The upper-left inset shows the top view of the HPM. The figure is not drawn to scale.
	}
	\label{fig:s3}
\end{figure}

Periodic boundary conditions are applied in all directions, and the system is coupled to a thermal bath. Directed transport is quantified by the net number of water molecules (NNWM) crossing the plane $z = 0$. A monotonic increase in NNWM indicates sustained vapor transport from interface B to interface A.

In a representative simulation (B1), the fixed charges are $Cl^-$ ions and the free ions are $Na^+$ ions. The system is simulated for 400 ns at 370 K. As shown in Supplementary Fig.~\ref{fig:s4}, NNWM exhibits a clear directional trend, demonstrating persistent transport. The transport cycle consists of evaporation at interface B, vapor migration through periodic boundaries, condensation at interface A, and liquid return flow from A to B. In contrast, a control simulation containing only the pure water slab and the HPM, obtained by removing all fixed and free ions (PW in Supplementary Fig.~\ref{fig:s4}), exhibits no directed flux, confirming the essential role of ion-induced asymmetry.

The primary advantage of this optimized model is that it prevents free ions from escaping from the liquid surface. However, it still requires nanoscale pore sizes in the HPM, making experimental implementation challenging.
\begin{figure}[H]
	\centering
	\includegraphics[width = 0.5\linewidth]{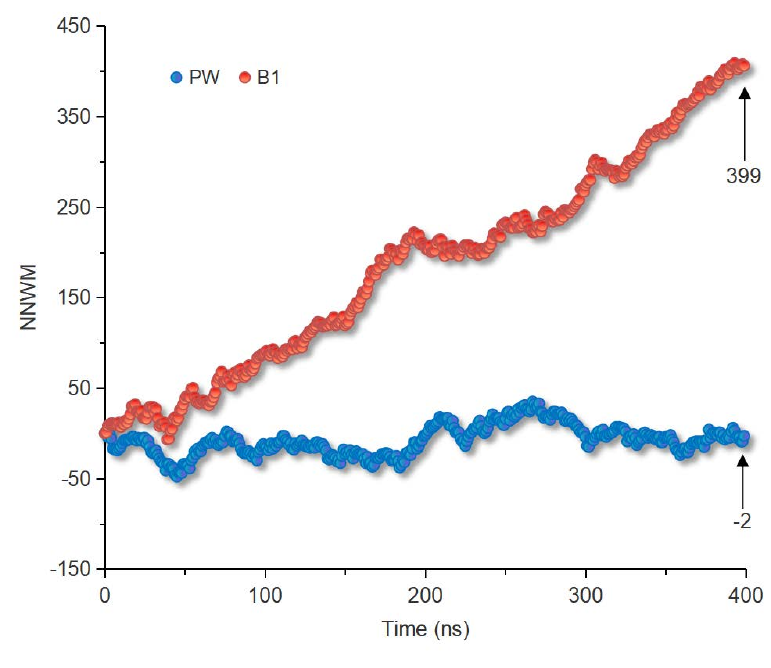}
	\caption{Variation of NNWM as a function of time for simulations PW and B1.
	}
	\label{fig:s4}
\end{figure}

\section{Additional Experimental Suggestions}\label{supp_sec:3}
As illustrated in Supplementary Fig.~\ref{fig:s5}, the partition and the HPM divide the water body into left and right liquid columns. A hydrophilic needle is positioned near the bottom of the right water column (above interface A), and only a single pore is present in the HPM. The distance between interfaces A and B is therefore approximately equal to the thickness of the HPM.
\begin{figure}[H]
	\centering
	\includegraphics[width = 0.5\linewidth]{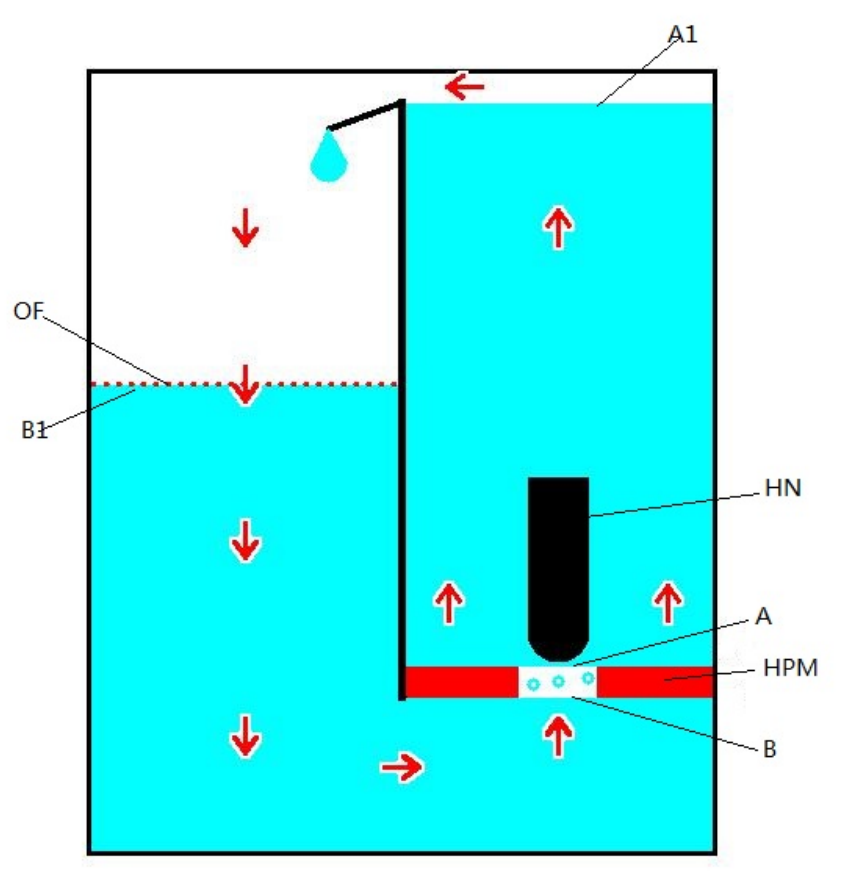}
	\caption{Schematic illustration of an experimental configuration. HPM, hydrophobic porous membrane; HN, hydrophilic needle; OF, oil film.
	}
	\label{fig:s5}
\end{figure}

This design offers several advantages:\\
1. When the needle compresses interface A, the validity of the model can be evaluated by observing whether a liquid-level difference develops between the two columns. If the model is correct, the liquid level A1 is expected to become higher than B1.\\
2. Weak flow rates can be converted into discrete droplets. When the generated pressure difference becomes sufficiently large that the liquid level A1 rises above the top of the partition, water will overflow to the other side. The flow rate can then be estimated by counting droplets. Furthermore, covering interface B1 with a thin oil film may improve measurement accuracy because the oil suppresses vapor–liquid exchange, making droplet transfer the only transport pathway.

Increasing the number of pores in the HPM can enhance the total flow rate. However, assigning one hydrophilic needle to each pore is experimentally difficult. An alternative design is shown in Supplementary Fig.~\ref{fig:s6}, in which each pore is paired with a hydrophilic particle with a surface modification layer of MPS. Each pore possesses a rough inner wall. The microscopic protrusions not only help immobilize the hydrophilic particle but also create gaps through which liquid can pass. Both theoretical and experimental studies have demonstrated that surface roughness can enhance hydrophobicity\cite{29,30}.
\begin{figure}[H]
	\centering
	\includegraphics[width = 0.5\linewidth]{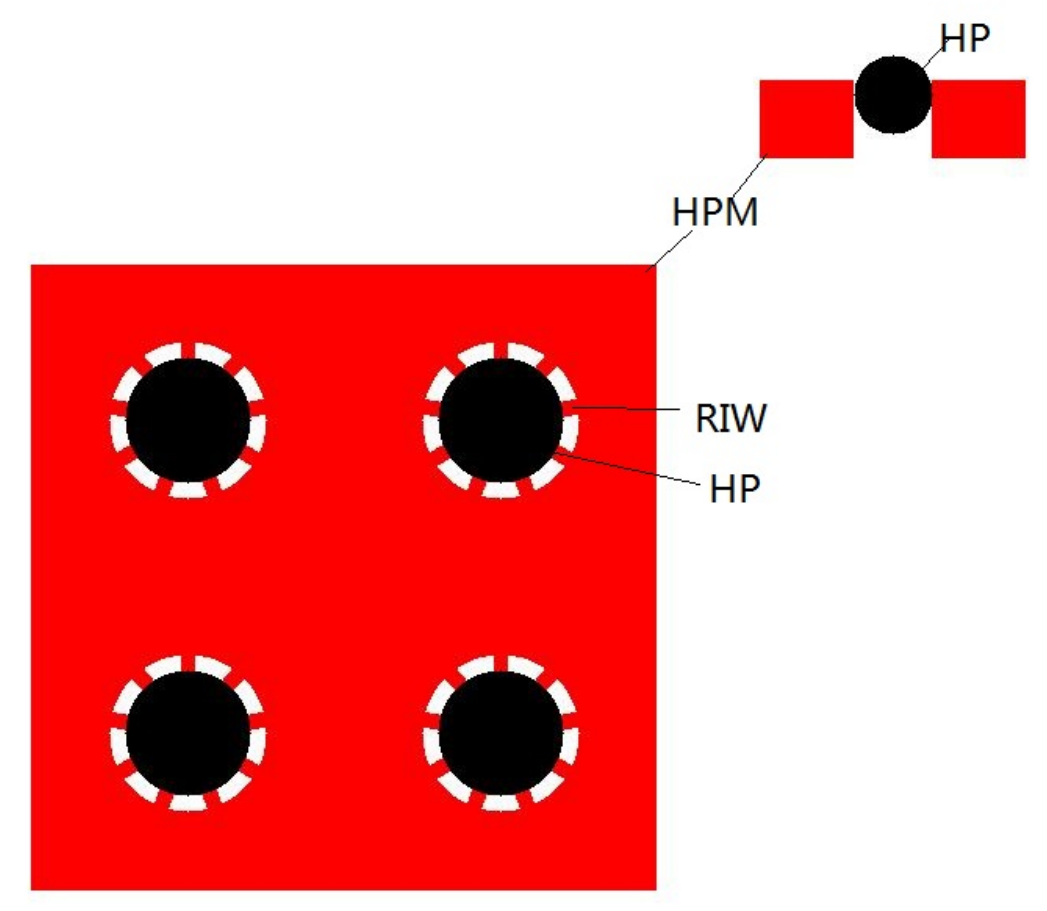}
	\caption{Schematic illustration of an HPM with multiple pores. HPM, hydrophobic porous membrane; HP, hydrophilic particle; RIW, rough inner wall.
	}
	\label{fig:s6}
\end{figure}

It should be noted that the top of each hydrophilic particle must protrude above the HPM surface so that it can contact the liquid. Once contact occurs, the liquid spontaneously wets the entire particle surface. Although this design does not allow flexible adjustment of the compression force in the same way as a movable needle, it offers the advantages of structural simplicity, low cost, and potentially higher flow rates.

\section{Flow Rate Estimation}\label{supp_sec:4}
For the system shown in Supplementary Fig.~\ref{fig:s5}, how many water droplets correspond to the daily flow rate? It should be noted that the liquid-level difference between A1 and B1 generates an additional pressure on the interfacial layer A:
\begin{equation}
	P_h = h \rho g
	\label{equ:s1}
\end{equation}
where $h$ is the liquid-level difference between the two interfaces (in this case, it’s taken as 0.01 m), $\rho$ is the density of water (taken as 1000 $kg m^{-3}$), and $g$ is the gravitational acceleration (taken as 9.8 $m s^{-2}$). The effect of $P_h$ on the chemical potential of water molecules at interface A is opposite to that of $P_n$. Therefore, the effective pressure responsible for reducing the chemical potential is $P_n - P_h$, which can be substituted into Eq. (11) of the main text to calculate the saturated vapor pressure of interface A. Assuming that the fluid behaves as an ideal fluid, the velocity of vapor transport through the pore driven by the saturated vapor pressure difference between the two interfaces can be calculated using Bernoulli’s equation. The following table presents the theoretical maximum single-pore flow rates for different pore radii at $T = 300 K$. Here, the saturated vapor pressure of interface B (pure water) is approximately $P_b = 3550$ Pa.

\begin{table}[htbp]
	\centering
	\setlength{\tabcolsep}{12pt}   
	\renewcommand{\thetable}{S1}
	\caption{Theoretical maximum flow rates corresponding to different pore radii.}
	\label{tab:s1}
	\begin{tabular}{lcccccc}
		\toprule
		$r_c(\mu m)$ & $P_n(Pa)$ & $P_a(Pa)$ & $\Delta P(Pa)$ & $v(ms^{-1})$ & $m_d(kg day^{-1})$  \\
		\midrule
		50    & 2,880    & 3,549.93    & 0.0715    & 2.37    & 4.10E-05     \\
		5     & 28,800    & 3,549.26   & 0.7381    &  7.6    & 1.32E-06   \\
		0.5    & 288,000    & 3,542.60     & 7.3964    & 24.06    & 4.17E-08     \\
		0.05    & 2,880,000    & 3,476.70    & 73.297    & 75.75   & 1.31E-09    \\
		\bottomrule
	\end{tabular}
\end{table}
$r_c(\mu m)$: radius of the pore (also the radius of curvature of the deformed liquid interface)\\
$P_n(Pa)$: pressure applied by the needle tip to interface A\\
$P_a(Pa)$: saturated vapor pressure of interface A\\
$\Delta P(Pa)$: saturated vapor pressure difference between interfaces A and B\\
$v(ms^{-1})$: vapor flow velocity\\
$m_d(kg day^{-1})$: mass flow rate of vapor, which also represents the amount of water condensed at interface A per day\\

As shown in the table, smaller pores generate larger saturated vapor pressure differences, whereas larger pores are favorable for achieving higher flow rates. When the pore radius is $50 \mu m$, the mass flow rate reaches $4.1 \times 10^{-5} kg day^{-1}$. Assuming that $1 g$ of water corresponds to approximately 30 drops, this flow rate is equivalent to 1.23 drops per day, indicating that experimental verification of the model by monitoring droplet formation is feasible.